\documentclass[12pt]{article}

\usepackage{latexsym}

\begin{document}

\newcommand{\ga}{\gamma}
\newcommand{\proof} { {\rm {\sc Proof.} } }
\newcommand{\qed}{\hfill$\Box$\par\medskip}
\newcommand{\qee}{\hfill$\Box$}

\newcommand{\hatl}{\hat L}
\newcommand{\D}{\hat D}
\newcommand{\X}{\hat X}
\newcommand{\Q}{{\bf Q}}

\newtheorem{thm}{Theorem}
\newtheorem{lemma}[thm]{Lemma}
\newtheorem{prop}[thm]{Proposition}

\title{Generalized Laplace transformations and integration of
hyperbolic systems of linear partial differential equations
\thanks{Submitted to
 ISSAC 2005, Beijing , China, July 24--27 2005.}
}
\author{
Sergey P.\ Tsarev\thanks{The research described in this article
was partially supported by  RFBR grant 04-01-00130.}
\\ Department of Mathematics  \\ Krasnoyarsk State Pedagogical
University \\ Lebedevoi 89, 660049 Krasnoyarsk, Russia \\
{\tt tsarev@newmail.ru}
 }

\date{January 14, 2005}

 \maketitle
\begin{abstract}
We give a new procedure for generalized factorization and
construction of the complete solution of strictly hyperbolic
linear partial differential equations or strictly hyperbolic
systems of such equations in the plane. This procedure generalizes
the classical theory of Laplace transformations of second-order
equations in the plane.
\end{abstract}

\section{Introduction} \label{SECT:intro}

Factorization of linear ordinary differential operators
\mbox{(LODOs)} is often used in modern algorithms for solution of
the corresponding differential equations. In the last 20 years
numerous modifications and generalizations of algorithms for
factorization of LODOs with rational function coefficients were
given (see e.g\ \cite{Bro1}). Such algorithms have close relations
with algorithms for computation of differential Galois groups and
closed-form (Liouvillian) solutions of linear ordinary
differential equations and systems of such equations (\cite{S-P}).
We have a nice and relatively simple theory of factorization of
LODOs.

For linear {\em partial} differential operators (LPDOs) and the
corresponding equations (LPDEs) the theory of factorization is
much more difficult. To the best of our knowledge there are only a
few theoretical results and only one algorithm for ``naive''
factorization of hyperbolic LPDO. In this introduction we will
give a brief account of the previously obtained results and state
our main result: existence of a recurrent procedure for
non-trivial factorization and finding closed-form complete
solutions of strictly hyperbolic systems of LPDEs in two
independent variables with coefficients in an arbitrary
differential field.

Theoretically one may propose several very different definitions
of factorization for LPDOs. The obvious ``naive'' definition
suggests to represent a given operator  $\hatl= \sum_{|\vec i|\leq
m} a_{i_1\cdots i_n}(\vec x) \D_{x_1}^{i_1}\D_{x_2}^{i_2}\cdots
\D_{x_n}^{i_n}$ as a composition of lower-order LODOs: $\hatl=
\hatl_1   \ldots \hatl_k$ with coefficients in some fixed
differential field. Unfortunately this definition does not enjoy
good theoretical properties: a given LPDO may have several very
different decompositions of this form, even the number of
irreducible factors $\hatl_s$ may be different, as the following
example (attributed in \cite{blum} to E.Landau) shows: if
\begin{equation}\label{landau}
  \begin{array}{c}
 \hat P=\D_x+x\D_y, \quad \hat Q=\D_x+1, \\
         \hat R=\D_x^2+x\D_x\D_y+\D_x+(2+x)\D_y,
\end{array}
\end{equation}
then  $\hat L = \hat Q   \hat Q   \hat P = \hat R   \hat Q$. On
the other hand the second-order operator $\hat R$ is absolutely
irreducible, i.e.\ one can not factor it into product of
first-order operators with coefficients in {\em any} extension of
$\Q(x,y)$. Still the ``naive'' definition of factorization may
help to solve the corresponding LPDE in some cases; recently
(\cite{grig2}) an algorithm for such factorization for the case of
hyperbolic LPDOs of arbitrary order was given.

In \cite{tsa,L-Sch-Ts} the adequate theoretical definition of
factorization and a factorization algorithm  for the case of
overdetermined systems with finite-dimensional solution space and
rational function coefficients  was given.

For a single second-order LPDO in two independent variables
\begin{equation}\label{2}
  \hatl = \D_x   \D_y -a(x,y)\D_x -b(x,y)\D_y -c(x,y)
\end{equation}
we have a very old and powerful theory of Laplace transformations
(not to be mixed with Laplace transforms!). We expose this nice
theory in Section~\ref{SECT:2}. Roughly speaking, an operator
(\ref{2}) is {\em Laplace-factorizable} if after several
applications of differential substitutions ($\D_x$- or
$\D_y$-transformations), which change the coefficients of
(\ref{2}) in a simple way, one obtains a naively-factorable
operator $\hatl_{(k)}=
(\D_y+b_{(k)})  (\D_x + a_{(k)})$ or $\hatl_{(k)}=(\D_x + a_{(k)})
(\D_y+b_{(k)})$.
This phenomenon of non-trivial Laplace-factorization explains the
existence of Landau example (\ref{landau}). The definition of
Laplace-factorizable operators turns out to be very fruitful in
applications, it was extensively used in classical differential
geometry (see e.g.\ \cite{dar-lec}) and actively studied in the
last decade in the framework of the theory of integrable nonlinear
partial differential equations \cite{anderson,sokolov,S-Zh}. This
is one of the most powerful methods of integration (construction
of the complete solution with the necessary number of functional
parameters) of the corresponding second-order equations in the
plane.

To the best of our knowledge the only serious effort to generalize
the classical theory of Laplace-factorization to operators of
higher order (in two independent variables) was undertaken in
\cite{LeRoux} with rather obscure exposition but deep insight and
a few enlightening remarks. Our approach, exposed in
Section~\ref{SECT:3}, gives a new uniform and general treatment of
this topic directly for $n\times n$ strictly hyperbolic systems in
two independent variables. Several modern papers
\cite{fer-lap,K-T1} investigate the theory of multidimensional
conjugate nets initiated in \cite[t.~4]{dar-lec}; this line of
research is in fact still in the domain of second-order equations
in two independent variables: the systems discussed in the cited
references are {\em overdetermined} systems with operators
(\ref{2}) and solution spaces parameterized by functions of one
variable. An interesting special case (operators (\ref{2}) with
matrix coefficients) was studied in \cite{S-Zh,Zh-St},
unfortunately the results are limited to this particular case of
higher-order systems.

A proper theoretical treatment of the factorization problem might
be expected in the framework of the $\cal D$-module theory (see
e.g\ \cite{cout} and a very good exposition of the appropriate
basic results in \cite{Lopat}). Unfortunately even in this modern
algebraic approach a ``good'' definition of factorization of LPDOs
with properties similar to the properties of factorization of
LODOs or commutative polynomials (decomposition of algebraic
varieties into irreducible components or primary decompositions in
Noetherian commutative rings) is not an easy task. Without going
into fine theoretical details we refer to \cite{ts98} where a
variant of such ``theoretically good'' definition of generalized
factorization of a single LPDO was given. As we have shown in
\cite{ts98}, this definition generalizes the classical theory of
Laplace-factorizable second-order operators. A drawback of this
theoretical approach was lack of any factorization algorithm for
a given LPDO.

In the present paper we give a new procedure (generalized Laplace
transformations) for generalized factorization and integration of
strictly hyperbolic LPDOs of arbitrary order with two independent
variables or systems of such LPDOs. Section~\ref{SECT:3} is
devoted to the detailed exposition of this new procedure.

In Section~\ref{SECT:4} we give an example of application of this
procedure to a $3\times3$ system and construct its complete
solution using the results of Section~\ref{SECT:3}. After this a
general scheme of generalized factorization and integration of a
strictly hyperbolic system in the plane is given. We conjecture
that this new procedure provides {\em an algorithm for generalized
factorization and closed-form complete solution} precisely in the
sense of \cite{ts98} if we limit the complexity of the answer.

\section{The classical heritage: Laplace transformations} \label{SECT:2}

Here we briefly sketch this classical theory
in a slightly different form suitable for our purpose. The
exhaustive exposition may be found in
\cite{dar-lec,forsyth,gour-l}. An arbitrary strictly hyperbolic
second-order equation with two independent variables $\hatl u =0$
and the operator
\begin{equation}\label{3}
  \hatl= \sum_{i=0}^2p_{i}\D_x^i\D_y^{2-i} + a_1(x,y)\D_x +
  a_2(x,y)\D_y + c(x,y),
\end{equation}
$p_i=p_{i}(x,y)$, may be rewritten in {\em characteristic form}
\begin{equation}\label{4}
  \begin{array}{l}
  (\X_1\X_2 + \alpha_1\X_1 + \alpha_2\X_2 +
  \alpha_3)u
  = \\
(\X_2\X_1 + \overline\alpha_1\X_1 + \overline\alpha_2\X_2 +
\alpha_3)u=0,
\end{array}
\end{equation}
where $\alpha_i=\alpha_i(x,y)$, the coefficients of the
first-order characteristic operators $\X_i = m_i(x,y)\D_x +
n_i(x,y)\D_y$ are found (up to a rescaling $\X_i \rightarrow
\gamma_i(x,y)\X_i$) from the characteristic equation
$m_i^2p_0-m_in_ip_1+n_i^2p_2=0$ for the principal symbol of
(\ref{3}). Since the operators $\X_i$ do not commute we have to
take into consideration in (\ref{4}) and everywhere below the {\em
commutation law}
\begin{equation}\label{cl}
  [\X_1,\X_2] = \X_1\X_2- \X_2\X_1 = P(x,y)\X_1 +Q(x,y)\X_2.
\end{equation}
Using the {\em Laplace invariants} of the operator (\ref{4}):
$$h=\X_1(\alpha_1) +\alpha_1\alpha_2 -\alpha_3, \quad 
k=\X_2(\overline\alpha_2) +\overline\alpha_1\overline\alpha_2-
\alpha_3,$$ we represent the original operator $\hatl$ in
partially factorized form
\begin{equation}\label{pf}
 \hatl = (\X_1+\alpha_2)  (\X_2+\alpha_1)-h=
  (\X_2+\overline\alpha_1) (\X_1+\overline\alpha_2) -k.
\end{equation}
From this form we see that the equation $\hatl u =0$ is equivalent
to any of the first-order systems
\begin{equation}\label{5}
(S_1):\left\lbrace  \begin{array}{l} \X_2u=-\alpha_1u +v,
\\  \X_1v =hu-\alpha_2v.
\end{array}\right. \Leftrightarrow
(S_2):\left\lbrace  \begin{array}{l}
 \X_1u=-\overline\alpha_2u +w, \\
  \X_2w =ku-\overline\alpha_1w.
\end{array}\right.
\end{equation}
\begin{prop}\label{prop1}
Any strictly hyperbolic LPDE  is equivalent to a $2\times2$
first-order characteristic system
\begin{equation}\label{6}
\left\lbrace  \begin{array}{l}
 \X_1u_1=\alpha_{11}(x,y)\,u_1 +\alpha_{12}(x,y)\,u_2, \\
  \X_2u_2 =\alpha_{21}(x,y)\,u_1 +\alpha_{22}(x,y)\,u_2,
\end{array}\right.
\end{equation}
with $\X_i = m_i(x,y)\D_x + n_i(x,y)\D_y$, $\X_1 \neq
\gamma(x,y)\X_2$, and any such system with non-diagonal matrix
$(\alpha_{ij})$ is equivalent to a second-order strictly
hyperbolic LPDE.
\end{prop}
\proof Transformation of a strictly hyperbolic LPDE into the form
(\ref{6}) is already given. The converse transformation is also
simple: if for example $\alpha_{12}\neq 0$ then substitute $u_2=
 (\X_1u_1 - \alpha_{11}u_1)/\alpha_{12}$ into the second equation
 of the system (\ref{6}). \qed
\begin{prop}\label{prop2}
If a $2\times2$ first-order system
\begin{equation}\label{7}
{\left(\!\! \begin{array}{l}
 v_1 \\ v_2
\end{array}\!\!\right)\!\!}_x  \!\!=
\left( \!\!\begin{array}{ll}
 a_{11} & a_{12} \\
  a_{21} & a_{22}\!\!\!\!
\end{array}\right)
{\left( \!\!\begin{array}{l}
 v_1 \\ v_2
\end{array}\!\!\right)\!\!}_y \!\!+
\left( \!\!\begin{array}{ll}
 b_{11} & b_{12} \\
  b_{21} & b_{22}
\end{array}\!\!\right)
\left( \!\!\begin{array}{l}
 v_1 \\ v_2
\end{array}\!\!\right)
\end{equation}
with $a_{ij}=a_{ij}(x,y)$, $b_{ij}=b_{ij}(x,y)$ is strictly
hyperbolic (i.e.\ the eigenvalues $\lambda_{k}(x,y)$ of the matrix
$(a_{ij})$ are real and distinct), then it may be transformed into
a system in characteristic form $(\ref{6})$.
\end{prop}
\proof Let $\lambda_{1}(x,y)$, $\lambda_{2}(x,y)$ be the
eigenvalues of $(a_{ij})$ and $\vec p_1 = (p_{11}(x,y),
p_{12}(x,y))$, $\vec p_2 = (p_{21}(x,y), p_{22}(x,y))$ be the
corresponding left eigenvectors: $\sum_k p_{ik}a_{kj} =
\lambda_ip_{ij}$. Form the operators $\X_i = \D_x - \lambda_i\D_y$
and the new characteristic functions \\ $u_i=\sum_kp_{ik}v_k$.
Then $\X_iu_i=\sum_k (\X_i p_{ik}) v_k
 + \sum_kp_{ik}\left( (v_k)_x - \lambda_i (v_k)_y\right)=
  \sum_{k,s}p_{ik} (a_{ks} - {} $ \\
  $\lambda_i \delta_{ks})(v_s)_y +
\sum_{k,s}p_{ik}b_{ks}v_s \! + \sum_k (\X_i p_{ik}) v_k
\!\!= \!\!
 \sum_s v_s \left(\sum_k p_{ik}b_{ks}\!+ (\X_i p_{is})\right)
 = \sum_k u_k \alpha_{ik}(x,y)$,
so we obtain the characteristic system (\ref{6}). \qed

The characteristic system (\ref{6}), equivalent to (\ref{7}), is
determined uniquely up to {\em operator rescaling} $\X_i
\rightarrow \gamma_i(x,y)\X_i$ and {\em gauge transformations}
$u_i \rightarrow g_i(x,y)u_i$. It is easy to check that the gauge
transformations to not change the Laplace invariants
 $h=
 \X_2(\alpha_{11}) - \X_1(\alpha_{22})
 -\X_1\X_2\ln(\alpha_{12})  -\X_1(P) +P\alpha_{11} +\alpha_{12}\alpha_{21}
 +(\alpha_{22}+\X_2(\ln\alpha_{12})+P)Q
 $ and
$k=\alpha_{12}\alpha_{21}$, they are just the Laplace invariants
of the operator (\ref{4}), obtained after elimination of $u_2$
from (\ref{6}). Rescaling transformations of $\X_i$ change them
multiplicatively $h \rightarrow \gamma_1\gamma_2 h$, $k
\rightarrow \gamma_1\gamma_2 k$. From the proofs we see that for a
{\em fixed} equation $\hatl u = 0$ with the operator (\ref{3}) we
obtain {\em two} different (inequivalent w.r.t.\ the scaling and
gauge transformations) characteristic systems (\ref{5}) and from
every fixed system (\ref{6}) we obtain {\em two} different
(inequivalent w.r.t.\ the gauge transformation $u \rightarrow
g(x,y)u$) hyperbolic LPDEs: one for the function $u_1$ and the
other for the function $u_2$. This observation gives rise to the
{\em Laplace cascade method of integration} of strictly hyperbolic
LPDEs $\hatl u=0$ with operators (\ref{3}):

$({\cal L}_1)$ If at least one of the Laplace invariants $h$ or
$k$ vanishes then the operator $\hatl$ factors (in the ``naive''
way) into composition of two first-order operators as we see from
(\ref{pf}); if we perform an appropriate change of coordinates
$(x,y) \rightarrow (\overline x, \overline y)$ (NOTE: for this we
have to solve  first-order nonlinear ODEs $dy/dx =
n_i(x,y)/m_i(x,y)$, cf.\ Appendix in \cite{grig2}) one can suppose
$\X_1=\D_{\overline x}$, $\X_2=\D_{\overline y}$ so we obtain the
complete solution of the original equation in quadratures: if for
example $\hatl u = (\D_{\overline x}+\alpha_2(\overline x,
\overline y))  (\D_{\overline y}+\alpha_1(\overline x, \overline
y))u =0$, then
 $$u= \exp\left(-\int \alpha_1 \, d\overline y \right) \left( X(\overline x)
  + \int Y(\overline y)
\exp\left(\int (\alpha_1\, d\overline y - \alpha_2\, d\overline
x)\right) d\overline y \right),$$
 where $ X(\overline x)$ and $
Y(\overline y)$ are two arbitrary functions of the characteristic
variables $\overline x$, $\overline y$.

$({\cal L}_2)$ If $h \neq 0$, $k \neq 0$, transform the equation
into one of the systems (\ref{5}) (to fix the notations we choose
the left system $(S_1)$) and then finding
\begin{equation}\label{difsub}
u = (\X_1 v +\alpha_2 v)/h
\end{equation}
substitute this expressions into the first equation of the left
system $(S_1)$ in (\ref{5}), obtaining a {\em $X_1$-transformed}
equation $\hatl_{(1)} v =0$. It has Laplace invariants
 (cf.\ \cite{anderson})
\begin{equation}\label{xl}
\begin{array}{l}
 h_{(1)}\!=\X_1(2\alpha_1\!-\!P)\! -\! \X_2(\alpha_2)\! -\!\X_1\X_2\ln h
 \!+\!Q\X_2\ln h  \! -\! \alpha_3\! +\!(\alpha_1\!-\! P)(\alpha_2\!-\!Q)
\!\!\!\!\!\!\!\!\!\!\!\!\!\! \\
 \ \ \ \  =\ \  2h-k - \X_1\X_2\ln h + Q\X_2\ln h +\X_2(Q)
-\X_1(P) +2PQ,\!\!\!\!\!
 \\[0.5em]
k_{(1)}=h.
\end{array}
\end{equation}
If $h_{(1)}=0$, we solve this new equation in quadratures and
using the same differential substitution (\ref{difsub}) we obtain
the complete solution of the original equation $\hatl u=0$.

$({\cal L}_3)$ If again $h_{(1)} \neq 0$, apply this
$X_1$-transformation several times, obtaining a sequence of
second-order operators $\hatl_{(2)}$, $\hatl_{(3)}$, \ldots\ of
the form (\ref{4}). If on any step we get $h_{(k)}=0$, we solve
the corresponding equation $\hatl_{(k)} u_{(k)} =0$ in quadratures
and, using the differential substitutions (\ref{difsub}), obtain
the complete solution of the original equation. Alternatively one
may perform {\em $\X_2$-transformations}: rewrite the original
equation in the form of the right system $(S_2)$ in (\ref{5}) and
using the substitution $u = (\X_2 w +\alpha_1 w)/k$ obtain the
equation $\hatl_{(-1)} w =0$ with Laplace invariants
\begin{equation}\label{yl}
\begin{array}{l}
h_{(-1)}=k, \\[0.5em]
 k_{(-1)}= 2k-h - \X_2\X_1\ln k - P\X_1\ln k
 +\X_2(Q)
  -\X_1(P) +2PQ.
\end{array}
\end{equation}
In fact this $\X_2$-transformation is a reverse of the
$\X_1$-trans\-for\-ma\-tion up to a gauge transformation (see
\cite{anderson}). So we have (infinite in general) chain of
second-order operators
\begin{equation}\label{ch}
   \ldots \stackrel{\X_2\!\!}{\leftarrow} \hatl_{(-2)} \stackrel{\X_2\!\!}{\leftarrow}
   \hatl_{(-1)}\stackrel{\X_2\!\!}{\leftarrow}   \hatl \stackrel{\X_1}{\rightarrow}
    \hatl_{(1)} \stackrel{\X_1}{\rightarrow} \hatl_{(2)} \stackrel{\X_1}{\rightarrow}
     \ldots
\end{equation}
and the corresponding chain of Laplace invariants
\begin{equation}\label{li}
   \ldots , h_{(-3)},\  h_{(-2)},\  h_{(-1)},\  h_0 = h,\
    h_{(1)},\ h_{(2)}, \ h_{(3)}, \ldots
\end{equation}
with recurrence formulas (\ref{xl}), (\ref{yl}). We do not need to
keep the invariants $k_{(i)}$ in (\ref{li}) since
$k_{(i)}=h_{(i-1)}$. If on any step we have $h_{(N)}=0$ then the
chains (\ref{ch}) and (\ref{li}) can not be continued: the
differential substitution (\ref{difsub}) is not defined; precisely
on this step the corresponding LPDE is trivially factorable and we
can find the complete solution for any of the operators of the
chain (\ref{ch}). For simplicity let us choose characteristic
variables $(\overline x, \overline y)$, so that
$\X_1=\D_{\overline x}$, $\X_2=\D_{\overline y}$. The complete
solution of the original equation in this case has the form
\begin{equation}\label{10}
\begin{array}{l}
\!\!\!\!\!\!  u = c_0(\overline x,\overline y)\left(F\!
   + \!\int\! G\beta \, d\overline y \right)
    + c_1(\overline x,\overline y) \left( F'\!+
      \!\int\! F\frac{\partial\beta}{\partial \overline x} \, d\overline y
    \right) +  \!\!\! \\[.5em]
 \ldots +
 c_n (\overline x,\overline y)\left( F^{(N)}\! + \!\int\! G
     \frac{\partial^N\beta}{\partial {\overline x}^N} \, d\overline y
     \right).
\end{array}
\end{equation}
where $F(\overline x)$, $G(\overline y)$ are two arbitrary
functions of the characteristic variables and $c_i(\overline
x,\overline y)$, $\beta(\overline x,\overline y)$ are some
definite functions obtained in the process of Laplace
transformations from the coefficients of the operator (\ref{3}).
 As one may prove (see e.g.\ \cite{dar-lec}) if the chain
(\ref{ch}) is finite in both directions (i.e.\ we have
$h_{(N)}=0$, $h_{(-K)}=0$ for some $N\geq 0$, $K\geq 0$) one may
obtain a quadrature-free expression of the general solution of the
original equation:
\begin{equation}\label{XY}
\!  u =\!\! c_0F + c_1F' + \!\ldots \!+ \! c_NF^{(N)}\!\! +
d_0\widetilde G + d_1\widetilde G' +
   \ldots +  d_{K+1}\widetilde G^{(K+1)}
\end{equation}
with definite $c_i(\overline x,\overline y)$, $d_i(\overline
x,\overline y)$ and  $F(\overline x)$, $\widetilde G(\overline y)$
--- two arbitrary functions of the characteristic variables and
vice versa: existence of ({\em a priori} not complete) solution of
the form (\ref{XY}) with arbitrary functions $F$, $G$ of
characteristic variables implies $h_{(s)}=0$, $h_{(-r)}=0$ for
some $s \leq N$, $r \leq K$. So {\em minimal differential
complexity} of the answer (\ref{XY}) (number of terms in it) is
equal to the number of steps necessary to obtain vanishing Laplace
invariants in the chains (\ref{ch}), (\ref{li}) and consequently
naively-factorable operators. Complete proofs of these statement
may be found in \cite[t.~2]{dar-lec}, \cite{forsyth,gour-l} for
the case $\X_1=\D_{x}$, $\X_2=\D_{y}$, for the general case cf.\
\cite[p. 30]{gour-l} and \cite{anderson}.

{\it Example 1.} As a straightforward computation shows, for the
equation $u_{xy} - \frac{n(n+1)}{(x+y)^2}u=0$  the chain
(\ref{li}) is symmetric ($h_{(i)} = h_{(-i-1)}$) and has length
$n$ in either direction. So the complexity of the answer
(\ref{XY}) may be very high and depends on some arithmetic
properties of the coefficients of the operator (\ref{3}); for the
equation $u_{xy} - \frac{c}{(x+y)^2}u=0$ the chains (\ref{ch}),
(\ref{li}) will be infinite unless the constant $c=n(n+1)$.

{\it Example 2.} For a stochastic ODE $\dot{x} = p(x)+\alpha(t)
q(x)$ with binary (dichotomic) noise $\alpha(t)=\pm 1$ and
switching frequency $\nu>0$ the averages $W(x,t)=\langle
\widetilde W(x,t)\rangle$ and $W_1(x,t)=\langle
\alpha(t)\widetilde W(x,t)\rangle$ for the probability density
$\widetilde W(x,t)$ in the space of possible trajectories $x(t)$
of the ODE satisfy a system of the form (\ref{7}) (see \cite{LL}):
\begin{equation}\label{WW}
   \left\lbrace  \begin{array}{l}
 W_t + \left(p(x)W\right)_x  + \left(q(x)W_1\right)_x =0, \\[0.5em]
 (W_1)_t + 2\nu W_1 +\left(p(x)W_1\right)_x  + \left(q(x)W\right)_x
 =0.
 \end{array}\right.
\end{equation}
The characteristic operators and left eigenvectors are simple:
$\X_i=\D_t-\lambda_i\D_x$, $\lambda_{1,2}=  -p(x) \pm q(x)$,
$p_{11}=p_{21}=p_{22}=1$, $p_{12}=-1$. The characteristic system
(\ref{6}) for the new characteristic functions $u_1=W-W_1$,
$u_2=W+W_1$ is
\begin{equation}\label{WWch}
   \left\lbrace  \begin{array}{l}
 \X_1u_1= -(p_x-q_x+\nu)\,u_1 +\nu \,u_2  ,\\
 \X_2u_2 = \nu \,u_1 - (p_x+q_x+\nu)\,u_2.
\end{array}\right.
\end{equation}
The Laplace invariants are
 $h= \nu^2 - [ p_{xx}q^2(p+q) + p_x^2q^2 - p_xq_xq(3p+q)
      -q_{xx}pq(p+q) q_x^2p(2p+q) ]/q^2 $,
 $k=\nu^2$, so if $\nu$,
$p(x)$ and $q(x)$ satisfy a second-order differential relation
$h=0$, one can solve (\ref{WW}) in quadratures. Especially simple
formulas may be obtained for polynomial $p(x)=p_1x+p_2x^2$,
$q(x)=q_2x^2$, $p_1>0$, $p_2<0$: in this case $k=\nu^2$,
$h=h_{(-2)}=\nu^2-p_1^2$ so if $\nu=p_1$ after the necessary
transformations we obtain the following quadrature-free expression
for the complete solution of the system (\ref{WW}):
 $$
\begin{array}{l}
\displaystyle W= \frac{q_2}{x^2}\left[ F'(\overline x)
-p_1F(\overline x) + p_1 G'(\overline y) - p_1^2 G(\overline y)
\right],\\[0.5em]
\begin{array}{l} \displaystyle  \!\! W_1=
\frac{1}{x^3}\left[ - q_2xF'(\overline x)
+p_1(p_2x+p_1)F(\overline x)
 + p_1 q_2 xG'(\overline y)   
 + p_1^2(p_2x+p_1) G(\overline y) \right],
\end{array}
\end{array}
 $$
where $\overline x = -t +\frac{1}{p_1}
\ln\frac{x}{p_1+(p_2+q_2)x}$, $\overline y = -t +\frac{1}{p_1}
\ln\frac{x}{p_1+(p_2-q_2)x}$ are the characteristic variables
($\X_2\overline x =0$, $\X_1\overline y =0$) and $F$, $G$ are two
arbitrary functions of the corresponding characteristic variables.
For the case $\nu^2 \!\neq \!p_1^2$ we can compute other Laplace
invariants of the chain (\ref{li}):
 $h_{(1)}\!=h_{(-3)}\!=\nu^2-4p_1^2$,
 $h_{(2)}=h_{(-4)}=\nu^2-9p_1^2$,
 $h_{(3)}=h_{(-5)}=\nu^2-16p_1^2$, \ldots
\ so for the fixed $p(x)=p_1x+p_2x^2$, $q(x)=q_2x^2$ and $\nu=\pm
p_1$, $\nu=\pm 2p_1$, $\nu=\pm 3p_1$, \ldots\ one can obtain
closed-form quadrature-free complete solution of the system
(\ref{WW}), with increasing complexity of the answer (\ref{XY}).

{\em Remark.} The forms (\ref{10}), (\ref{XY}) for the complete
solution are local: in the general case due to nontrivial
topological picture of the trajectories of the vector fields
$\X_1$, $\X_2$ in the plane we are unable to guarantee existence
of the global coordinate change $(x,y)\mapsto(\overline x,
\overline y)$ (cf.\ Example~2 above).

\section{Generalized Laplace transformations of
{\lowercase{ \Large $n\times n$}} hyperbolic systems}
\label{SECT:3}

Hereafter we suppose that the LPDE of order $n\geq 2$
\begin{equation}\label{3.1}
  \hatl u = \sum_{i+j\leq n} p_{i,j}(x,y)\D_x^i\D_y^j u =0
\end{equation}
is strictly hyperbolic, i.e.\ the characteristic equation
$\displaystyle \!\!\sum_{i+j=n}p_{i,j}\lambda^i=0$ has $n$ simple
real roots $\lambda_k(x,y)$.
\begin{prop}\label{l31}
Any strictly hyperbolic LPDE $(\ref{3.1})$ is equivalent to a
$n\times n$ first-order system in characteristic form
\begin{equation}\label{3.2}
 \X_iu_i= \sum_k \alpha_{ik}(x,y)u_k.
\end{equation}
\end{prop}
\proof The principal ($n$th-order) part of (\ref{3.1}) decomposes
into the product of the characteristic operators
$\X_i=\D_x-\lambda_i\D_y$ modulo lower-order terms. Other
lower-order terms of any given order $s$ may be also written as
sums of products of $\X_i$ (modulo terms of order $<s$) in the
following unique way:
\begin{equation}\label{XX}
\begin{array}{l} \displaystyle
\hatl = \X_1\X_2\cdots \X_n  
{}+\sum_{s=1}^{n-1} \sum_{i_1<\ldots < i_s\leq s+1} \!\!\! \!\!\!
a_{s,i_1\cdots i_s}(x,y)\X_{i_1}\cdots\X_{i_s}  \! +
a_0(x,y).\!\!\!
\end{array}
\end{equation}
This is easily proved using induction over $s$. For $s=1$ the
terms of order 1 are $p_{1,0}(x,y)\D_x +p_{0,1}(x,y)\D_y =
a_1(x,y)\X_1 + a_2(x,y)\X_2$ since the eigenvalues $\lambda_1$,
$\lambda_2$ are distinct. If this is proved for terms of order
$\leq s_1$ we may suppose (after a local change of variables) that
$\X_{s_1+2}=\D_x$ so
 $ \displaystyle \sum_{i+j=s_1+1} p_{i,j}\D_x^i\D_y^j = 
  p_{0,s_1+1}\D_y^{s_1\!+1}  +
  $ \\  $\displaystyle \left(\sum_{i+j=s_1}
   \!\! p_{i+1,j}\D_x^i\D_y^j\right)\!\D_x \! =
 q(x,y)\X_1\X_2\cdots \X_{s_1\!+1} 
  + \left(\sum_{i+j=s_1} \overline
p_{i+1,j}\D_x^i\D_y^j\right)\D_x = $ \\  $\displaystyle
 q(x,y)\X_1\X_2\cdots \X_{s_1+1} 
  +  \left(
\sum_{i_1<\ldots < i_{s_1}\leq s_1+1}
 \!\!\! \!\!\! a_{s_1,i_1\cdots i_{s_1}}
(x,y)\X_{i_1}\cdots\X_{i_{s_1}}\right)\X_{s_1+2} $ (modulo terms
of lower order).

Now, provided we have the form (\ref{XX}) of the operator $\hatl$,
introduce the characteristic functions
 $u_n\equiv u$, $u_{n-1}=\X_nu_n+\alpha_{nn}u_n$,
 $u_{n-2}=\X_{n-1}u_{n-1}+\alpha_{n-1,n-1}u_{n-1}+\alpha_{n-1,n}u_n$,
 \ldots , $u_1 = \X_2u_2 +\sum_{s=2}^n\alpha_{2s}u_s$
with so far undefined $\alpha_{pq}=\alpha_{pq}(x,y)$. We have to
define them inductively so that the expression $\hat M = \X_1u_1
+\sum_{s=1}^n\alpha_{1s}u_s$ (again with $\alpha_{1s}(x,y)$ to be
determined) would coincide with the operator $\hatl$. In fact, the
highest-order term in $\hat M$ is just $\X_1\X_2\cdots \X_n$. The
terms of order $n-1$ in $\hat M$ are just $n$ products
$\X_{i_1}\cdots\X_{i_{n-1}}$, $i_1<\ldots <i_{n-1}\leq n$ (in
every such product exactly one of $\X_i$, $1\leq i \leq n$, is
missing) with coefficients $\alpha_{kk}$, $1\leq k\leq n$, so we
just fix $\alpha_{kk}$ to be equal the respective $
a_{n-1,i_1\cdots i_{n-1}}$ in (\ref{XX}). This makes $\hat M -
\hatl$ to be of order $n-2$. Each coefficient at the terms
$\X_{i_1}\cdots\X_{i_{n-2}}$ of order $n-2$ in $\hat M - \hatl$
includes derivatives of the already defined $\alpha_{kk}$ and
precisely one of $\alpha_{k-1,k}$, $2\leq k\leq n$, so choosing
the latter appropriately we lower the order of $\hat M - \hatl$
again. This step-down process will give us all $\alpha_{pq}$ so we
obtain the following system, equivalent to $\hatl u=0$:
$$   \left\lbrace  \begin{array}{l}
 \X_1u_1 = -\sum_{s=1}^n\alpha_{1s}u_s, \\
 \X_2u_2 = -\sum_{s=2}^n\alpha_{2s}u_s +u_1, \\
 \ldots \\
 \X_nu_n = -\alpha_{nn}u_n +u_{n-1}. \\
\end{array}\right.
$$
 \qed

{\em Remark 1}. There are many ways of transforming a given
higher-order LPDE into a first-order system. But for our purposes
we need a {\em precise equivalence}, guaranteeing a one-to-one
correspondence between the solution space of the original equation
and the solution space of the system. If for example we take the
equation $u_{xx}-u_{yy}-a(x,y)u=0$ and (after
\cite[Ch.~I]{Courant}) form the following first-order system
\begin{equation}\label{K3}
\left\lbrace   \begin{array}{l}
   u_x=p, \\ q_x=p_y,\\ p_x=q_y+a(x,y)u,
\end{array}\right.
\end{equation}
then we shall fix for it limited Cauchy data $q|_{x=0}=u_y|_{x=0}$
(so taking some {\em subset} of solutions of (\ref{K3})) in order
to have a one-to-one correspondence between the solution space of
the original equation and some subset of the solution space of
(\ref{K3}). Another straightforward possibility
\begin{equation}\label{K3p}
\left\lbrace   \begin{array}{l}
   u_x=p, \\ u_y=q,\\ p_x=q_y+a(x,y)u
\end{array}\right.
\end{equation}
gives in fact an overdetermined degenerate system, it can not be
put into the standard form (\ref{3.3}) below.

{\em Remark 2}.
 The converse of Proposition~\ref{l31} is not true:
already for $n=3$ a generic system (\ref{3.2}) gives  for each
$u_i$ an overdetermined involutive system of equations (see for
example \cite[Ch.~I]{Courant}). So hereafter we will construct our
generalized Laplace transformations directly for systems in
characteristic form (\ref{3.2}) with $\X_i \neq \mu(x,y)\X_j$ for
$i \neq j$. But for $n \times n$ first-order systems the
equivalence can be proved. This is a well-known fact (see for
example \cite{Courant}), we give its simple proof for completeness
of our exposition.
\begin{prop}\label{l32}
Any $n\times n$ first-order linear system
\begin{equation}\label{3.3}
(v_i)_x = \sum_{k=1}^n a_{ik}(x,y)(v_k)_y +
  \sum_{k=1}^n b_{ik}(x,y)v_k
\end{equation}
with strictly hyperbolic matrix $(a_{ik})$ (i.e.\ with real and
distinct eigenvalues of this matrix) is equivalent to a system in
characteristic form $(\ref{3.2})$.
\end{prop}
\proof A \ straightforward \ calculation \ similar \ to \ that \
in the \ proof \ of \ Proposition~\ref{prop2} \ shows \ that \ if
\
 $\vec p_i = {}$ \\$(p_{i1}(x,y), \ldots ,
p_{in}(x,y))$ are the left eigenvectors, $\sum_k p_{ik}a_{kj} =
\lambda_ip_{ij}$, $\X_i = \D_x - \lambda_i\D_y$ are the
characteristic operators, then the new characteristic functions
$u_i=\sum_kp_{ik}v_k$ satisfy a system of the form (\ref{3.2}).
\qed

The characteristic form (\ref{3.2}) of a given system (\ref{3.3})
is defined up to {\em operator rescalings} $\X_i \rightarrow
\gamma_i(x,y)\X_i$ and {\em gauge transformations} $u_i
\rightarrow g_i(x,y)u_i$. The quantities
$h_{ij}=\alpha_{ij}\alpha_{ji}$,
$h_{ijk}=\alpha_{ij}\alpha_{jk}\alpha_{ki}$, $i \neq j \neq k$,
and other similar cyclic products of the coefficients of
(\ref{3.2}) are invariant w.r.t.\ the gauge transformations so it
is natural to call them {\em Laplace invariants} of (\ref{3.2}).
The complete set of such invariants can be found using the
standard methods \cite{Olver,Athorne}. We will not dwell upon this
problem.

Our generalization of the Laplace transformations is based on the
following observation.
For a given $2\times 2$ system
(\ref{6}) the procedure of $\X_1$-transformation consists of the
following steps:
$$\hatl u=0 \mapsto
 \left\lbrace  \begin{array}{l} \X_2u=-\alpha_1u +v \\
 \X_1v =hu-\alpha_2v
\end{array}\right.
 \mapsto
 \hatl_{(1)}v=0.
$$
If we will try to reformulate this in terms of 
$2\times 2$ characteristic systems (\ref{6}), we arrive at the
following procedure:

$({\cal L}_1)$ For a given system (\ref{6}) choose one of the two
variables $u_i$ and eliminate it, for example we may eliminate
$u_2$ using the first equation of (\ref{6}): $u_2 = (\X_1 u_1
-\alpha_{11}u_1)/\alpha_{12}$. Substitute it into the other
equation of the system obtaining a second-order equation for
$u_1$:
 $\X_2 (\frac{1}{\alpha_{12}}(\X_1u_1 - \alpha_{11}u_1)) - \alpha_{21}u_1
  - \frac{\alpha_{22}}{\alpha_{12}}(\X_1u_1 - \alpha_{11}u_1) =
  0$.

$({\cal L}_2)$ Swap the operators $\X_1$ and $\X_2$ in the
second-order term of this equation using the appropriate
commutation laws and then rearrange the first-order terms
obtaining an equal expression of the form
 $\X_1 \left(\frac{1}{\alpha_{12}}(\X_2u_1)\right) +p_1(x,y)\X_1u_1  +p_2(x,y)\X_2u_1
 +p_3(x,y)u_1  =
 \X_1 \left(\frac{1}{\alpha_{12}}(\X_2u_1 + {p_{1}}{\alpha_{12}}u_1)\right)
 + \frac{\overline p_2}{\alpha_{12}}(\X_2u_1 + {p_{1}}{\alpha_{12}}u_1) +
 \overline p_3(x,y)u_1  =0$.

$({\cal L}_3)$ The last form of this second-order equation
immediately suggests to set $\overline u_2 = (\X_2u_1 +
{p_{1}}{\alpha_{12}}u_1)/\alpha_{12}$ so we can rewrite the
equation in the form of the transformed $2\times 2$ characteristic
system
\begin{equation}\label{nlsys}
 \left\lbrace  \begin{array}{l}
 \X_2u_1=-p_1\alpha_{12} u_1 +\alpha_{12}\overline u_2, \\
  \X_1\overline u_2 =-p_2 u_1 -\overline  p_3(x,y)\overline u_2.
\end{array}\right.
\end{equation}
{\em Remark}. One should not be mistaken: if we apply this
``swapping procedure'' twice we {\em do not necessarily return} to
the original system. This depends on the choice of the
characteristic function to eliminate on step $({\cal L}_1)$ in
(\ref{nlsys}): eliminating $\overline u_2$ we certainly return to
the original system (\ref{6}), but eliminating $u_1$ from
(\ref{nlsys}) we actually move one more step forward in the chain
(\ref{ch}).

 This commutation trick gives the basis for our {\em
generalized Laplace transformations}:

$({\cal L}_1)$ For a given $n\times n$ characteristic system
(\ref{3.2}) choose one of its equations with a non-vanishing
off-diagonal coefficient  $\alpha_{ik}\neq 0$, find
 $u_k = (\X_iu_i - \sum_{s \neq k} \alpha_{is}u_s)/\alpha_{ik}$
and substitute this expression into all other equations of the
system. We obtain one second-order equation
\begin{equation}\label{L11}
   \X_k \big(\frac{1}{\!\alpha_{ik}\!}(\X_iu_i - \!\sum_{s \neq k}\alpha_{is}u_s)\big)
    - \!\sum_{p\neq k}\alpha_{kp}u_p
  - \frac{\alpha_{kk}}{\alpha_{ik}}(\X_iu_i - \!\sum_{s \neq k}\alpha_{is}u_s)\! =\!
  0
\end{equation}
and $n-2$ first-order equations
\begin{equation}\label{L12}
  \X_j u_j -\sum_{s \neq k}\alpha_{js}u_s -
  \frac{\alpha_{jk}}{\alpha_{ik}}(\X_iu_i - \sum_{s \neq k}\alpha_{is}u_s) =
  0,
\quad j \neq i,k.
\end{equation}
$({\cal L}_2)$  The second step consists in rewriting the system
(\ref{L11}), (\ref{L12}) in the following form with slightly
modified unknown functions $\overline u_j = u_j + \rho_j(x,y)u_i$,
$j \neq i,k$, $\overline u_i \equiv u_i$, new coefficients
$\beta_{pq}(x,y)$ but the same characteristic operators $\X_p$:
\begin{equation}\label{L21}
\!\!  \X_i (\frac{1}{\alpha_{ik}\!\!}(\X_k\overline u_i
   - \!\sum_{s \neq k}\beta_{is}\overline u_s))
    - \!\sum_{p\neq k}\beta_{kp}\overline u_p
  - \frac{\beta_{kk}\!}{\alpha_{ik}\!}(\X_k\overline u_i -
    \!\sum_{s \neq k}\beta_{is}\overline u_s) \!=\!
  0,
\end{equation}
\begin{equation}\label{L22}
 \X_j \overline u_j -\sum_{s \neq k}\beta_{js}\overline u_s -
  \frac{\beta_{jk}}{\alpha_{ik}}(\X_k\overline u_i - \sum_{s \neq k}\beta_{is}\overline u_s) =
  0, \quad j \neq i,k.
\end{equation}
As we prove below this is always possible in a unique way.


$({\cal L}_3)$  Introducing $\overline u_k
=\frac{1}{\alpha_{ik}}(\X_k\overline u_i - \sum_{s \neq
k}\beta_{is}\overline u_s)$ rewrite (\ref{L21}), (\ref{L22}) as
the transformed characteristic system
\begin{equation}\label{L31}
      \left\lbrace  \begin{array}{l}
 \X_i \overline u_k =  \sum_{p}\beta_{kp}\overline u_p ,  \\[0.5em]
 \X_k\overline u_i = \sum_{s \neq k}\beta_{is}\overline u_s
  +\alpha_{ik}\overline u_k, \\[0.5em]
 \X_j \overline u_j =\sum_{s}\beta_{js}\overline u_s,
  \quad j \neq i,k.
 \end{array}\right.
\end{equation}

{\em Correctness of the step $({\cal L}_2)$}. First of all we
rewrite the l.h.s.\ of (\ref{L11}) (again using the appropriate
commutation laws) as $E_i = \X_i\big(\frac{1}{\alpha_{ik}}(\X_k
u_i)\big) + p_i\X_iu_i + p_k\X_ku_i -
 \sum_{s \neq i,k}\frac{\alpha_{is}}{\alpha_{ik}}\X_ku_s
  +\sum_{s \neq k}c_{is}u_s$
and the l.h.s.\ of (\ref{L12}) as $E_j = \X_j u_j -\sum_{s \neq
k}c_{js} u_s -   p_{jk}\X_ku_i-   p_{jj}\X_ju_i $, where we have
used the fact that any 3 characteristic operators are linearly
dependent (we have only 2 independent variables!): $\X_i =
\phi_j(x,y)\X_k+\psi_j(x,y)\X_j$. Introducing new $\overline u_j =
u_j - p_{jj}(x,y)u_i$, $j \neq i,k$,  $\overline u_i \equiv u_i$
 we have
$E_j = \X_j \overline u_j  -\sum_{s \neq k}\overline c_{js}
\overline u_s - p_{jk}\X_k \overline u_i$  and  $E_i
=\X_i\big(\frac{1}{\alpha_{ik}}(\X_k \overline u_i)\big) +
p_i\X_i\overline u_i + {{\overline p}}_{k}\X_ku_i - \sum_{s \neq
i,k}\frac{\alpha_{is}}{\alpha_{ik}}\X_k\overline u_s +\sum_{s \neq
k}\overline c_{is}\overline u_s =
\X_i\big(\frac{1}{\alpha_{ik}}(\X_k \overline u_i)\big) +
p_i\X_i\overline u_i + {\overline p}_{k}\X_k\overline u_i - 
 \sum_{s \neq i,k}\frac{\alpha_{is}}{\phi_s\alpha_{ik}}
 (\X_i\overline u_s - \psi_s\X_s\overline
 u_s)
  +\sum_{s \neq k}\overline c_{is}\overline u_s $. 
So adding to it a suitable combination of $E_j$ we obtain \\
 $\overline E_i= E_i -
 \sum_{s \neq i,k}\frac{\psi_s\alpha_{is}}{\phi_s\alpha_{ik}} E_s=
 \X_i\big(\frac{1}{\alpha_{ik}}
 (\X_k \overline u_i + \alpha_{ik}p_i\overline u_i
   - \sum_{s \neq i,k}\frac{\alpha_{is}}{\phi_s}\overline
    u_s
 )\big)
 + \big(\frac{{\overline {\overline p}}_{k}}{\alpha_{ik}}(\X_k \overline u_i
 + \alpha_{ik}p_i\overline u_i
   - \sum_{s \neq i,k}\frac{\alpha_{is}}{\phi_s}\overline
    u_s
)\big) +  
\sum_{s \neq k}{\overline{\overline c}}_{is}\overline u_s$,
which coincides with (\ref{L21}), the basic coefficients
$\beta_{ks}(x,y)$, $\rho_i(x,y)$ are therefore defined. Now
(\ref{L22}) is in fact the equation $E_j = 0$ with slightly
rearranged terms. \qed

\section{An example and the general factorization procedure} \label{SECT:4}

Let us consider the following $3\times 3$ system
\begin{equation}\label{3x3}
   \left\lbrace  \begin{array}{l}
   \X_1u_1= u_1 +2u_2 +u_3, \\
   \X_2u_2 = -6u_1 +u_2 +2u_3, \\
   \X_3u_3 = 12u_1 +6u_2+u_3,
\end{array}\right.
\end{equation}
with $\X_1=\D_x$, $\X_2=\D_y$, $\X_3=\D_x+\D_y$. Let us eliminate
$u_3$ using the first equation:
\begin{equation}\label{eu3}
  u_3 = \X_1u_1-  u_1 -2u_2.
\end{equation}
Substituting this expression into the other equations we obtain
the following two equations:
$$ E_2=\X_2u_2 +6u_1 -u_2 - 2(\X_1u_1-  u_1 -2u_2) =0,
$$
$$ E_3 = \X_3(\X_1u_1-  u_1 -2u_2) -12 u_1 -6u_2 - (\X_1u_1-  u_1 -2u_2)
  =0.
$$
This completes the step $({\cal L}_1)$  of our generalized Laplace
transformation.

Step $({\cal L}_2)$ is done in 4 substeps:

1) Change  $\X_1u_1$ to $\X_3u_1-\X_2u_1$ in $E_2$ and collect the
terms with $\X_2$: $E_2=\X_2u_2 +6u_1 -u_2 - 2\X_1u_1 + 2u_1 +4u_2
=\X_2u_2 +8u_1 +3u_2 - 2\X_3u_1 + 2\X_2u_1 =
 \X_2(u_2 + 2u_1) +8u_1 +3u_2 - 2\X_3u_1 = 0$,
so we now can introduce the new function $\overline  u_2 =u_2 +
2u_1$. Change everywhere $u_2$ to $\overline  u_2 -2u_1$: \\
 $ E_2=\X_2\overline  u_2 +2u_1
+3\overline  u_2 - 2\X_3u_1 =0,$ \\
$ E_3 = \X_3\X_1u_1 +3  \X_3u_1 -2\X_3\overline  u_2
   -3 u_1 -4\overline  u_2 - \X_1u_1  =0.
$

2) Swap $\X_3$ and $\X_1$ in $E_3$, substitute
 $\X_3\overline  u_2=\X_1\overline  u_2+\X_2\overline  u_2$ and collect the terms with
 $\X_1$: \\
$ E_3 = \X_1 (\X_3u_1 -u_1 -2\overline  u_2 )
 +3  \X_3u_1 -2\X_2\overline  u_2
   -3 u_1 -4\overline  u_2  =0.
$
 This suggests to set $\overline  u_3=(\X_3u_1 -u_1 -2\overline  u_2 )$.

3) Change everywhere $\X_3u_1$ to $\overline  u_3+u_1 +2\overline
u_2 $:
\\
$
E_2=\X_2\overline  u_2 +2u_1 +3\overline  u_2 - 2(\overline
u_3+u_1 +2\overline  u_2) =
 \X_2\overline  u_2  - \overline  u_2 - 2\overline  u_3 = 0, $\\
 $E_3 = \X_1 \overline  u_3 +3 (\overline  u_3+u_1 +2\overline  u_2) -2\X_2\overline  u_2
   -3 u_1 -4\overline  u_2  =
    \X_1 \overline  u_3  +3 \overline  u_3  -2\X_2\overline  u_2
    +2\overline  u_2  =0.
$

4) Now get rid of the term  $-2\X_2\overline  u_2$ in $E_3$: \\ $
\overline  E_3 =E_3 +2E_2 = \X_1 \overline  u_3 -\overline  u_3
  =0.
$\\ The resulting equations $E_2= \X_2\overline  u_2  - \overline
u_2 - 2\overline  u_3 = 0$,
 $\overline  E_3 = \X_1 \overline u_3
-\overline  u_3=0$ and $\overline  u_3=(\X_3u_1 -u_1 -2\overline
u_2 )$ give us the transformed system:
\begin{equation}\label{tr3x3}
   \left\lbrace  \begin{array}{l}
   \X_1 \overline u_3=  \overline u_3, \\
   \X_2\overline u_2 =  2 \overline u_3 + \overline u_2 , \\
   \X_3u_1 = \overline u_3 +2 \overline u_2+ u_1.
\end{array}\right.
\end{equation}
Its equations can be consecutively solved: the first equations
gives $\overline u_3= e^x F(y)$. Using the standard method of the
variation of constants we solve the second equation obtaining
$
\overline u_2 = e^y (G(x) +2\int e^{x-y}F(y)\,dy) $. In order to
remove the quadrature we introduce a new arbitrary function
$\widetilde F(y)$ instead of $F(y)$: $F(y)=e^y\widetilde F'(y)$.
Integrating by parts we obtain
 $\overline u_2 = e^y (G(x) +2 e^{x}\widetilde F(y))$.
The last equation gives $u_1 = C(x,y)\exp\frac{x+y}{2}$ where
$C(x,y)$ satisfies the equation
 $\X_3(C)=\exp\frac{-x-y}{2}(\overline u_3 +2\overline u_2)=
  \exp\frac{x+y}{2}(\widetilde F'(y) + 4 \widetilde F(y)) + 2 \exp\frac{y-x}{2} G(x)$.
 Again introduce new arbitrary functions $\widetilde G(x)$,
$\overline F(y)$, such that
 $G(x)=\widetilde G'(x)$,
 $\widetilde F(y) = e^{-y}\overline F'(y)$, then we can
 find
$C(x,y)=2 \exp\frac{y-x}{2}\widetilde  G(x) +
 \exp\frac{x-y}{2}( 3\overline F(y) + \overline F'(y)) + H(x-y)$.
Finally we get the following complete solution of the transformed
system (\ref{tr3x3}):
$$ \!   \left\lbrace  \begin{array}{l}
 u_1 =  
2e^y\widetilde G(x) + e^{x}(3 \overline F(y) + \overline F'(y)) +
 \exp\frac{x+y}{2}H(x-y),
 \\[0.5em]
 \overline u_2 = e^y\widetilde G'(x)
 + 2e^{x}\overline F'(y)
  , \!\!\!\!\!\! \\[0.5em]
 \overline u_3 =e^{x}(\overline F''(y) - \overline F'(y)
  ),
\end{array}\right.
$$
where $\overline F(y)$, $\widetilde G(x)$ and $H(x-y)$ are three
arbitrary functions of one variable each. The complete solution of
the original system (\ref{3x3}) can be easily found using the
differential substitutions of the steps  $({\cal L}_1)$ and
$({\cal L}_2)$: $u_2 = \overline u_2 - 2u_1$,
 $u_3=\X_1u_1-u_1-2u_2=\X_1u_1+3u_1-2\overline u_2$
(and certainly $u_1\equiv u_1$!).

{\bf \ \ \ The general factorization procedure}.

I) If we have to solve an equation $\hatl u=0$ or a system
(\ref{3.3}) transform it into the characteristic form (\ref{3.2}).

II) If the matrix $(\alpha_{ij}(x,y))$ of the characteristic
system is upper- or lower-triangular (similar to the matrix of
(\ref{tr3x3})) solve the equations consecutively.

III) If the matrix is block-triangular, the system {\em factors}
into several lower-order systems; try for each subsystem the step
IV.

IV) In the general case of not-triangular matrix
$(\alpha_{ij}(x,y))$ perform several (consecutive) generalized
Laplace transformations, using different choices of the {\em pivot
element} $\alpha_{ik}\neq 0$ on step  $({\cal L}_1)$. The goal is
to obtain a block-triangular matrix for one of the transformed
systems.

The main problem of this procedure is absence of upper bounds for
the number of steps. As we have seen in Section~\ref{SECT:2} ({\em
Example~1}) this bound depends on the coefficients of the equation
(system) in a nontrivial way. On the other hand in this situation
(the classical Laplace-factorization and our generalization) the
procedure of transformation {\em does not require solution of
differential equations} and {\em does not depend on the
differential field of coefficients}, contrary to the case of
LODOs.

\section*{Acknowledgements} We would like to thank
Prof.~V.M.~Loginov who provided many interesting examples of
hyperbolic systems originating in the theory of stochastic
differential equations and Prof.~D.~Grigoriev, Prof.~F.~Schwarz
for stimulating discussions in the long period of this research
started with the paper \cite{ts98}.

\end{document}